# Point Group Symmetry and Deformation Induced Symmetry Breaking of Superlattice Materials

Pu Zhang[1], Albert C. To[2]

Department of Mechanical Engineering and Materials Science, University of Pittsburgh, Pittsburgh, PA 15261, USA

**Abstract:** The point group symmetry of materials is closely related to their physical properties and quite important for material modeling. However, superlattice materials have more complex symmetry conditions than crystals due to their multilevel structural feature. Thus, a theoretical framework is proposed to characterize and determine the point group symmetry of nonmagnetic superlattice materials systematically. A variety of examples are presented to show the symmetry features of superlattice materials in different dimensions and scales. In addition, the deformation induced symmetry breaking phenomenon is also studied for superlattice materials, which has potential application in tuning physical properties by imposing a strain field.

---

[1] Email: p_zhang87@hotmail.com
[2] Corresponding author. Email: albertto@pitt.edu

# 1 Introduction

Superlattice materials are lattice materials comprising two or more structural levels, which exhibit distinct features compared to single crystals. For example, the unit cell of superlattice materials ranges from nanometers to millimeters. In addition, the material distribution in the unit cell of superlattice materials could be designed in favor of the required performance. Therefore, superlattice materials exhibit prominent physical properties and multifunctional performance that are usually unattainable in single crystals. The applications of superlattice materials are quite broad, which include, but are not limited to, cellular materials [1-4], phononic/photonic crystals [5-10], materials with controllable heat/electron conductivity [11-13], periodic metamaterials [14-16], nanoparticle superlattice materials [17, 18], and two-phase periodic composites [19, 20]. Therefore, superlattice materials have drawn much attention by researchers from fabrication to application. Particularly, the design and fabrication of superlattice materials have benefited from the modern fabrication techniques developed in the past decades. For example, the widely used techniques include material removal method (e.g. etching), additive manufacturing [1-3], physical or chemical deposition [13], and self-assembly [17]. Up to now, superlattice materials of different material types have been fabricated from 1D to 3D and at all scales including nano-, micro-, and macro-scale.

Similar to crystals, the long-range physical properties of superlattice materials must be compatible with their point group symmetry according to the Neumann's law [21]. The relations between point group symmetry and physical properties of crystals have been studied extensively and outlined in [21, 22]. The physical properties (e.g. elastic, thermal, dielectric, optic, etc.) are usually represented by different ranks of tensors, which are invariant or form-invariant under symmetry transformations of the material point group. Therefore, the long-range physical properties of superlattice materials could be determined qualitatively once their symmetry is known. However, the point group symmetry of superlattice materials is more complex than that of single crystals. Note that the unit cell of a superlattice material may have complex topology and material heterogeneity that cannot be easily determined through visualization. Thus, there is a strong demand for a theoretical framework to analyze and determine the point group symmetry of superlattice materials systematically, which is one primary aim of this work.

Another interesting problem is the deformation induced symmetry breaking [23-26], which arises in superlattice materials (or single crystals as well) once a strain field is imposed. This phenomenon is important in at least two applications. First, the symmetry breaking of superlattice materials may lead to changes of their physical properties, which opens opportunities to tune/control the physical performance and design functional materials. For example, deformation induced acoustic, optical, and thermal property changes in materials are reported in the literature [24, 27-29]. Secondly, the symmetry evolution is also important for the constitutive modeling of superlattice materials because the symmetry property changes must be considered when modeling the material property evolution during deformation [30, 31]. Therefore, the symmetry evolution of superlattice materials after deformation is discussed in this work with an emphasis on the symmetry breaking at small strain cases.

The main goal of this work is to study the point group symmetry characteristics of superlattice materials. To achieve this goal, a theoretical framework is proposed to describe and determine the point group symmetry of superlattice materials with a wide range of examples outlined and discussed. Thereafter, the symmetry breaking of deformed superlattice materials are introduced. In addition, special remarks on the relation between symmetry and physical properties of superlattice materials are addressed at the end. Note that the analysis is for nonmagnetic superlattice materials.

## 2   Symmetry of Superlattice Materials

Two examples are illustrated in Fig. 1 to show the multilevel structural feature of superlattice materials and the complexity of their symmetry. Different from single crystals, superlattice materials have symmetry and translation order at multiple structural levels, and each level may have different symmetry properties. For example, Figure 1(a) shows a cubic lattice material at level 1 (the coarsest scale) with its constituent material exhibiting a hexagonal lattice at level 2 (finer scale). In contrast, Fig. 1(b) illustrates a nanocrystal superlattice [32] with FCC structures at both level 1 and level 2, but with different lattice orientations. In summary, the symmetry of superlattice materials has three unique properties compared with that of single crystals. (1) The point group symmetry of superlattice may not be the same at different levels (e.g. Fig. 1(a)). (2) The lattice orientations may be distinct across different levels, even if they belong to the same point group (e.g. Fig. 1(b)). (3) The material components of the superlattice unit cell could have different materials, orientations, and symmetries. It is thus clear that the multilevel structural feature and complex unit cell topology lead to great difficulty in determining the overall symmetry of superlattice materials by visualization alone. To address this issue, a theoretical framework is established to characterize and determine the point group symmetry of superlattice materials in a systematic manner. Without loss of generality, we focus on superlattice materials with two structural levels since they are commonly seen. The proposed theory could be easily applied to superlattice materials with three levels or more [33, 34] using a hierarchical approach.

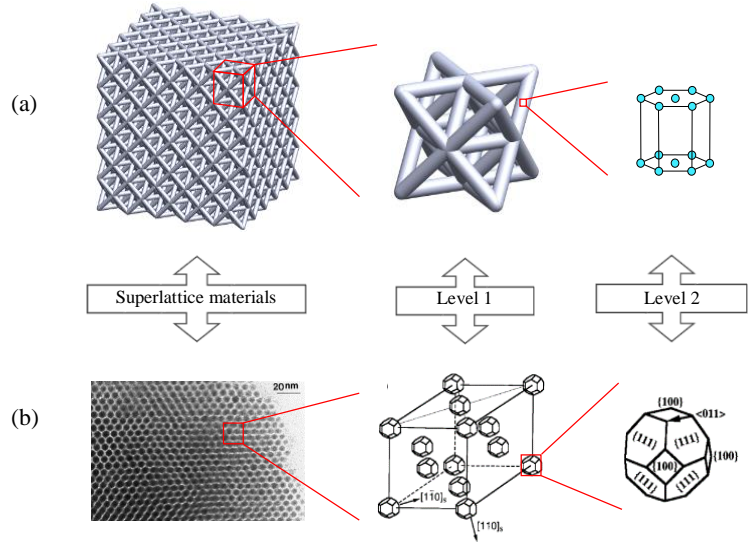

**FIGURE 1.** Two typical examples of superlattice materials and their multilevel structural features. (a) A cubic cellular material composed of a material with hexagonal lattice in level 2. All the ligaments have the same material type and orientation. (b) Self-assembled nanocrystal superlattice with FCC structures in both levels. However, the orientations of the lattices are different at the two levels. (Images in (b) are reprinted with permission from [32]. ©1996 American Chemical Society)

The notations used in this work are introduced. We use $\mathbb{R}$ for the real number set, $\mathbb{N}$ for the natural number set, $\mathbb{R}^d$ for the $d$-dimension Euclidean space ($d$ = 2 or 3), lowercase Greek letters for scalars (e.g. $\alpha, \beta$), lowercase bold-face Latin letters for vectors (e.g. $\mathbf{a}, \mathbf{b}$), uppercase bold-face Latin letters for tensors of rank 2 (e.g. $\mathbf{A}, \mathbf{B}$) and rank $\geqslant 3$ (e.g. $\mathbf{A}, \mathbf{B}$), and uppercase bold-face script letters for point groups (e.g. $\boldsymbol{\mathcal{A}}, \boldsymbol{\mathcal{B}}$). The inner products are defined as $\mathbf{A}\mathbf{b} = A_{ij}b_j\mathbf{e}_i$ and $\mathbf{A}\mathbf{B} = A_{ij}B_{jk}\mathbf{e}_i \otimes \mathbf{e}_k$, where $\mathbf{e}_i$ is the basis vector, $\otimes$ is the dyadic product, and Einstein's summation rule applies. Some other tensor operations will be defined in the text. In addition, the Hermann-Mauguin notation [35] is adopted to represent the material point groups.

The point group of a two-level superlattice material is determined by examining its symmetry at both level 1 and level 2. Thus, the symmetry of the topology, material type, material orientation, and local material point group should be evaluated when the reference configuration $\mathcal{B} \subset \mathbb{R}^d$ is mapped to the transformed configuration $\mathcal{B}' \subset \mathbb{R}^d$ (see Fig. 2), where $\mathcal{B}$ and $\mathcal{B}'$ are sets containing all material points and the prime symbol indicates quantities or fields after a symmetry transformation. Hence, a material point $\mathbf{X} \in \mathcal{B}$ will be mapped to the point $\mathbf{X}' \in \mathcal{B}'$ after the transformation [36]. The symmetry of superlattice materials is more complicated than the classical point group theory of single crystals, which only considers the symmetry of the atom location and species in a unit cell. Therefore, we will divide the symmetries of superlattice materials into two categories: Topology symmetry and material symmetry.

(1) *Topology symmetry*. This includes the symmetries of the geometry and material type, which are determined at level 1. Herein the material type merely means the material name and phase (e.g. copper with FCC lattice), regardless of the material orientation. The material type field is characterized by a scalar function $\varphi(\mathbf{X}) : \mathbb{R}^d \mapsto \mathbb{N}$ since the material type of each material point could be labeled as a natural number. Thereafter, the topology point group $\mathcal{T} = \{\mathbf{T}\}$ of the superlattice material is defined as

$$\mathcal{T} := \{\mathbf{T} \in \mathcal{O}(d) \mid \varphi(\mathbf{T}\mathbf{X}) = \varphi(\mathbf{X}), \forall \mathbf{X} \in \mathcal{B}\} \tag{1}$$

where $\mathcal{O}(d)$ represents the orthogonal group in $\mathbb{R}^d$. Equation (1) is derived from the fact that $\varphi' \equiv \varphi(\mathbf{X}') = \varphi$ when the map $\mathbf{X} \mapsto \mathbf{X}' = \mathbf{T}\mathbf{X}$ does not change the material type field. Alternatively, the topology symmetry of the superlattice material is preserved once the material type field $\varphi(\mathbf{X})$ is invariant under a symmetry transformation.

(2) *Material symmetry*. This includes the symmetry properties of the local material orientations and local point groups at level 2 of a superlattice material. Generally, not all operations in the topology point group $\mathcal{T}$ guarantee the material symmetry at level 2. Hence, by denoting the overall point group of the superlattice material as $\mathcal{G} = \{\mathbf{G}\}$, it is obvious that $\mathcal{G}$ must be a subgroup of $\mathcal{T}$, as

$$\mathcal{G} \leq \mathcal{T} \tag{2}$$

A concept of material point group field is introduced first, i.e. $\mathcal{M}(\mathbf{X})$, which is a function mapping an arbitrary material point $\forall \mathbf{X} \in \mathcal{B}$ to its corresponding local material point group. Note that $\mathcal{M}(\mathbf{X})$ could even have different orders for different material points. As shown in Fig. 2, the material point group field $\mathcal{M}(\mathbf{X})$ should be form-invariant under symmetry transformations, as

$$\mathcal{M}' \equiv \mathcal{M}(\mathbf{X}') = \mathbf{G}\mathcal{M}(\mathbf{X})\mathbf{G}^{\mathrm{T}}, \forall \mathbf{X} \in \mathscr{B} \tag{3}$$

where the superscript 'T' indicates the transpose operation and $\mathbf{G}\mathcal{M}\mathbf{G}^{\mathrm{T}} := \{\mathbf{GMG}^{\mathrm{T}} \mid \forall \mathbf{M} \in \mathcal{M}\}$. Therefore, the point group $\mathcal{G}$ is defined based on Eqs. (2) and (3), as

$$\mathcal{G} := \{\mathbf{G} \in \mathcal{T} \mid \mathcal{M}(\mathbf{GX}) = \mathbf{G}\mathcal{M}(\mathbf{X})\mathbf{G}^{\mathrm{T}}, \forall \mathbf{X} \in \mathscr{B}\} \tag{4}$$

The point group of single crystals can be described solely by Eq. (1), the invariant of a scalar field $\varphi(\mathbf{X})$, whereas that of superlattice materials also requires the form-invariant of a tensor group field $\mathcal{M}(\mathbf{X})$ shown in Eq. (4). Generally speaking, Eqs. (1) and (4) provide all information required to determine the point group of a superlattice material. However, it is usually inefficient and impractical to exhaust the material type and point group invariant at each material point. Therefore, a simpler method will be introduced in Section 3.

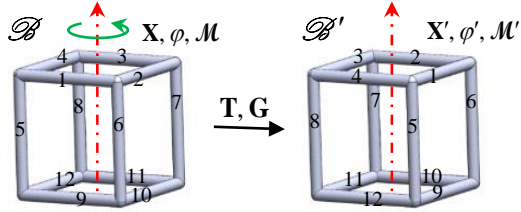

**FIGURE 2.** Schematic illustration of the reference configuration $\mathscr{B}$ and its transformed configuration $\mathscr{B}'$ for a cubic superlattice material. The material is rotated for 90° as an example of the symmetry transformation. The field quantities $\varphi$ and $\mathcal{M}$ are transformed to $\varphi'$ and $\mathcal{M}'$, respectively, under the symmetry transformation $\mathbf{T}$ or $\mathbf{G}$.

## 3  Determination of Point Group of Superlattice Materials

### 3.1  Overview of the Method

A practical method is proposed to determine the point group of the superlattice material. The superlattice unit cell is divided into $n_M$ material components, each with the same material type and orientation, and then their symmetry properties are compared. For example, the unit cell in Fig. 2 has twelve material components ($n_M = 12$), which are numbered in sequence. Denote the symmetry orders of the point groups $\mathcal{G}$ and $\mathcal{T}$ as $n_G$ and $n_T$, respectively, which satisfy $n_G \leq n_T$ according to Eq. (2). Thus, under the *l*-th topology symmetry transformation $\mathbf{T}_l \in \mathcal{T}$ ($l \leq n_T$) at level 1, a material component is transformed to the position of another one, forming a pair of material components whose local material symmetries at level 2 should be examined. Consequently, a total of $n_l = n_M$ (or $n_l = n_M/2$ for some cases) pairs of material components will be formed under the *l*-th transformation $\mathbf{T}_l \in \mathcal{T}$. Finally, the superlattice point group $\mathcal{G}$ will be determined after evaluating the material symmetries of all $\sum_{l=1}^{n_T} n_l$ pairs of material components for every topology symmetry transformation in $\mathcal{T}$.

## 3.2 Symmetry of a Pair of Material Components

As illustrated in Fig. 3, the material symmetry of two material components $\mathcal{B}_{lk}$ and $\mathcal{B}'_{lk}$ is studied first, which are the *k*-th pair ($k \leq n_l$) of material components related to the topology symmetry transformation $\mathbf{T}_l \in \mathcal{T}$, i.e. the configuration transformation is $\mathbf{X} \mapsto \mathbf{X}' = \mathbf{T}_l \mathbf{X}$ for $\forall \mathbf{X} \in \mathcal{B}_{lk}$. For the sake of convenience, two local Cartesian coordinate systems $\mathbf{x}_{lk}$ and $\mathbf{x}'_{lk}$ are established for the material components $\mathcal{B}_{lk}$ and $\mathcal{B}'_{lk}$, respectively, according to the convention of crystallography for lattice orientations. By this means, the material point groups of $\mathcal{B}_{lk}$ and $\mathcal{B}'_{lk}$ have a unified form as $\bar{\mathcal{M}}_{lk} = \{\bar{\mathbf{M}}_{lk}\}$ in their own local coordinate systems, which brings about great simplification for the analysis. A brief introduction to the default lattice orientation and material point groups is attached in Appendix A. Note that the material point groups of $\mathcal{B}_{lk}$ and $\mathcal{B}'_{lk}$ can also be represented in the global coordinate system $\mathbf{X}$, which is correlated to the local coordinate systems by $\mathbf{x}_{lk} = \mathbf{Q}_{lk} \mathbf{X}$ and $\mathbf{x}'_{lk} = \mathbf{Q}'_{lk} \mathbf{X}$, where $\mathbf{Q}_{lk}$ and $\mathbf{Q}'_{lk}$ are orthogonal transformation matrices. Therefore, in the global coordinate system, the material point groups $\mathcal{M}_{lk}$ and $\mathcal{M}'_{lk}$ of the two material components $\mathcal{B}_{lk}$ and $\mathcal{B}'_{lk}$ are expressed as

$$\begin{aligned} \mathcal{M}_{lk} &= \mathbf{Q}_{lk}^{\mathrm{T}} \bar{\mathcal{M}}_{lk} \mathbf{Q}_{lk} \\ \mathcal{M}'_{lk} &= \mathbf{Q}'^{\mathrm{T}}_{lk} \bar{\mathcal{M}}_{lk} \mathbf{Q}'_{lk} \end{aligned} \qquad (5)$$

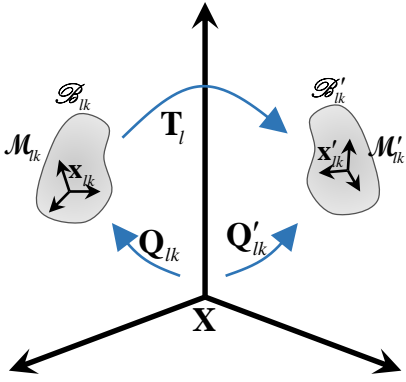

**FIGURE 3.** Schematic illustration of the material symmetry transformation of a pair of material components in a superlattice unit cell.

Further, it is obtained from Eq. (5) that the relation between $\mathcal{M}_{lk}$ and $\mathcal{M}'_{lk}$ is

$$\mathcal{M}'_{lk} = \mathbf{Q}'^{\mathrm{T}}_{lk} \mathbf{Q}_{lk} \mathcal{M}_{lk} \mathbf{Q}^{\mathrm{T}}_{lk} \mathbf{Q}'_{lk} \tag{6}$$

Equation (6) implies that $\mathcal{M}_{lk}$ and $\mathcal{M}'_{lk}$ are conjugate [37].

The material symmetry of $\mathcal{B}_{lk}$ and $\mathcal{B}'_{lk}$ is examined under the corresponding topology symmetry transformation $\mathbf{T}_l \in \mathcal{T}$. Based on Eq. (4), the material symmetry of $\mathcal{B}_{lk}$ and $\mathcal{B}'_{lk}$ requires that $\mathbf{T}_l \mathcal{M}_{lk} \mathbf{T}^{\mathrm{T}}_l = \mathcal{M}'_{lk}$. In addition, the point group $\mathcal{M}'_{lk}$ does not change under a symmetry transformation of its own members, i.e. $\mathcal{M}'_{lk} = \mathbf{M}'_{lk} \mathcal{M}'_{lk} \mathbf{M}'^{\mathrm{T}}_{lk}$. Finally, it is obtained that the material symmetry condition of the two material components is

$$\begin{aligned}
\mathbf{T}_l \mathcal{M}_{lk} \mathbf{T}^{\mathrm{T}}_l &= \mathcal{M}'_{lk} \\
&= \mathbf{M}'_{lk} \mathcal{M}'_{lk} \mathbf{M}'^{\mathrm{T}}_{lk} \\
&= \mathbf{Q}'^{\mathrm{T}}_{lk} \mathbf{Q}_{lk} \mathbf{M}_{lk} \mathcal{M}_{lk} \mathbf{M}^{\mathrm{T}}_{lk} \mathbf{Q}^{\mathrm{T}}_{lk} \mathbf{Q}'_{lk}, \quad \text{for } \forall \mathbf{M}_{lk} \in \mathcal{M}_{lk}
\end{aligned} \tag{7}$$

where the last equality is derived by using Eq. (6). Thus it is found from Eq. (7) that the material symmetry of the *k*-th pair of material components can only be conserved when

$$\mathbf{T}_l \in \mathbf{Q}'^{\mathrm{T}}_{lk} \mathbf{Q}_{lk} \mathcal{M}_{lk} \tag{8}$$

Finally, the symmetry condition for the pair of material components is obtained by substituting Eq. (5) into Eq. (8), as

$$\mathbf{T}_l \in \mathbf{Q}'^{\mathrm{T}}_{lk} \bar{\mathcal{M}}_{lk} \mathbf{Q}_{lk} \tag{9}$$

## 3.3 Point Group of Superlattice Materials

As mentioned in Section 3.1, the overall point group $\mathcal{G}$ of a superlattice material can be obtained by examining the material symmetry of all material component pairs for each $\mathbf{T}_l \in \mathcal{T}$. Therefore, as a direct generalization to Eq. (9), $\mathbf{T}_l$ will satisfy the material symmetry of all its $n_l$ material component pairs if and only if

$$\mathbf{T}_l \in \bigcap_{k=1}^{n_l} \mathbf{Q}_{lk}'^{\mathrm{T}} \bar{\mathcal{M}}_{lk} \mathbf{Q}_{lk} \tag{10}$$

Furthermore, the overall point group $\mathcal{G}$ is determined by examining the condition in Eq. (10) for $\forall \mathbf{T}_l \in \mathcal{T}$, as

$$\mathcal{G} := \left\{ \mathbf{T}_l \mid \mathbf{T}_l \in \mathcal{T} \text{ and } \mathbf{T}_l \in \bigcap_{k=1}^{n_l} \mathbf{Q}_{lk}'^{\mathrm{T}} \bar{\mathcal{M}}_{lk} \mathbf{Q}_{lk} \right\} \tag{11}$$

Equation (11) expresses the general form of the overall point group $\mathcal{G}$ for a superlattice material, which is straightforward to evaluate by computation. Some remarks are noted when applying Eq. (11). First, it is suggested that one evaluates $\mathcal{G}$ in Eq. (11) by starting from the symmetry transformation generators $\mathbf{T}^g \in \mathcal{T}$ [35], which may reduce a great amount of work. Moreover, in case that one symmetry transformation $\mathbf{T}$ is excluded from $\mathcal{G}$, several similar ones can also be excluded by comparing the subgroups [35] of $\mathcal{T}$.

Even though the general form of $\mathcal{G}$ is shown in Eq. (11), several special cases are worthwhile to be introduced in particular, which are quite useful and applicable to most superlattice materials found in the literature. There are at least four special cases as follows.

(C1) $\mathcal{G} = \mathcal{T} \cap \bar{\mathcal{M}}$. This is achieved when $\mathbf{Q}_{lk} = \mathbf{Q}_{lk}' = \mathbf{I}$ and $\bar{\mathcal{M}}_{lk} = \bar{\mathcal{M}}$, namely, all material components have the same orientations and material point groups. This case is quite useful for many superlattice materials, e.g. the one in Fig. 1(a).

(C2) $\mathcal{G} = \mathcal{T} \cap \mathbf{Q}^{\mathrm{T}} \bar{\mathcal{M}} \mathbf{Q}$. This is a generalization to the case C1 when $\mathbf{Q}_{lk} = \mathbf{Q}_{lk}' = \mathbf{Q}$ and $\bar{\mathcal{M}}_{lk} = \bar{\mathcal{M}}$, i.e. the local coordinate system is disoriented with the global one. A typical example is the nanocrystal superlattice material in Fig. 1(b).

(C3) $\mathcal{G} = \mathcal{T}$. Surprisingly, the overall symmetry of the superlattice can still be identical to its topology symmetry, which is achieved when $\mathbf{T}_l \in \mathbf{Q}'^{\mathrm{T}}_{lk} \bar{\mathcal{M}}_{lk} \mathbf{Q}_{lk}$ for all pairs of material components. This also provides a chance that a material with low-order symmetry can be carefully arranged to design a superlattice material with high-order overall symmetry. In addition, $\mathcal{G} = \mathcal{T}$ is always valid if the constituent material is isotropic.

(C4) $\mathcal{G} = \{\mathbf{I}\}$. This is a case with the least order symmetry. Note that this happens frequently for superlattice materials since the topology symmetry and material symmetry cannot always be guaranteed simultaneously.

## 4 Superlattice Materials in Different Dimensions

Superlattice materials have been synthesized or fabricated from 1D to 3D. A variety of representative examples are introduced in this section to show the symmetry features of superlattice materials in different dimensions. In addition, the symmetry of particles will also be studied as an additional application of the theory.

### 4.1 3D Superlattice Materials

Typical 3D superlattice materials include the cellular materials, nanocrystal superlattices, periodic two-phase composites, and among others. Generally speaking, the point groups of these 3D superlattice materials are within the 32 crystal point groups and 7 continuous groups, whose symmetry transformations are briefly introduced in Appendix A. Some examples are shown to further introduce the symmetry of 3D superlattice materials.

The cubic cellular structure in Fig. 1(a) is studied first. It is obvious that the topology point group $\mathcal{T}$ in level 1 is $m\bar{3}m$ with an order of 48. In contrast, the material point group $\bar{\mathcal{M}}$ in level 2 is $6/mmm$ with an order of 24. In addition, the local and global coordinate systems coincide with each other, namely, $\mathbf{Q}_{lk} = \mathbf{Q}'_{lk} = \mathbf{I}$. Therefore, this is exactly the special case C1, and the point group of this superlattice material is $\mathcal{G} = \mathcal{T} \cap \bar{\mathcal{M}}$. After comparing the transformations in these two groups, it is found that $\mathcal{G}$ is the point group $mmm$ with an order of 8. Thus, the overall symmetry of this cubic cellular material is in the orthorhombic class, quite different from the topology symmetry and material symmetry.

The nanocrystal superlattice in Fig. 1(b) shows different feature compared with the cubic cellular structure in Fig. 1(a). In this case, both the topology symmetry and material symmetry belong to the $m\bar{3}m$ group since they are both FCC structures. Hence, the overall point group $\mathcal{G}$ would also be $m\bar{3}m$ should the lattice orientations coincide in the two levels. However, it is not the case in Fig. 1(b). According to Wang [18], the orientation correlation between the superlattice (level 1) and nanocrystal lattice (level 2) is $[110]\|[110]_s$ and $[001]\|[1\bar{1}0]_s$ (implying $[1\bar{1}0]\|[00\bar{1}]_s$), where the subscript "s" indicates the Miller index of the nanocrystal. Therefore, it is derived that $\mathbf{Q}_{lk} = \mathbf{Q}'_{lk} = \mathbf{Q}$ with

$$\mathbf{Q} = \begin{bmatrix} \frac{1}{2} & \frac{1}{2} & \frac{1}{\sqrt{2}} \\ \frac{1}{2} & \frac{1}{2} & \frac{-1}{\sqrt{2}} \\ \frac{-1}{\sqrt{2}} & \frac{1}{\sqrt{2}} & 0 \end{bmatrix} \tag{12}$$

This nanocrystal superlattice is a typical example of the special case C2, whose overall point group is determined by $\mathcal{G} = \mathcal{T} \cap \mathbf{Q}^T \bar{\mathcal{M}} \mathbf{Q}$. It is finally found that $\mathcal{G}$ belongs to the orthorhombic point group *mmm* with the three reflection planes as $(110)$, $(1\bar{1}0)$, and $(001)$ of the superlattice (see Appendix B for details). Note that the obtained point group $\mathcal{G}$ is a conjugate group of the *mmm* group listed in the table of point group transformations [38, 39].

So far, the two examples discussed above are for the crystal point groups. Actually the proposed theoretical framework can also be applied to the continuous point groups (see Appendix A). Consider the cubic lattice structure shown in Fig. 2 with a topology point group $\mathcal{T}$ of $m\bar{3}m$. Assume that the constituent material is transversely isotropic (e.g., $\bar{\mathcal{M}}$ is $\infty/mm$) with the privileged axis along the [001] direction of the superlattice, as shown in Fig. 4(a). In this case, the overall point group $\mathcal{G}$ is found to be $\bar{4}/mmm$ of the tetragonal class. However, note that the overall symmetry is affected by the orientation of the privileged axis of the constituent material. For instance, if the privileged axis is rotated to the [111] direction (see Fig. 4(b)), the overall symmetry $\mathcal{G}$ of the cellular structure will be $\bar{3}m$ of the trigonal class with the three-fold axis along the [111] direction, which is totally different from the one in Fig. 4(a). This analysis has particular application to 3D printed superlattice materials [1, 2] since the constituent material often shows transversely isotropic behavior with the privileged axis along the print orientation. Therefore, the 3D printing direction will definitely affect the overall symmetry behavior and also the physical/mechanical properties [22] consequently.

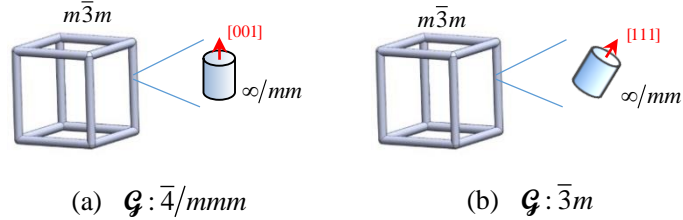

(a) $\mathcal{G}:\bar{4}/mmm$  (b) $\mathcal{G}:\bar{3}m$

**FIGURE 4.** Point group symmetry of a cubic cellular material with transversely isotropic materials. (a) The privileged axis of the material is along the [001] direction of the superlattice. The overall point group is $\bar{4}/mmm$. (b) The privileged axis of the material is along the [111] direction of the superlattice. The overall point group is $\bar{3}m$.

### 4.2   2D Superlattice Materials

The 2D superlattice materials have a 2D lattice structure at the superlattice level (level 1), which include 2D cellular materials, 2D phononic/photonic crystals, graphene nanomesh [12], etc. Similar to the 3D material point groups, there are a total of 10 crystal point groups and two continuous point groups in 2D. Note that all of the transverse isotropy groups degenerate to the isotropic case in 2D. Although the symmetry analysis to 2D superlattice materials is similar to the 3D cases in Section 4.1, there are still some unique features. Two examples are studied to show the symmetry characteristics of the 2D superlattice materials.

Strictly speaking, a 2D superlattice material requires both the superlattice (level 1) and material lattice (level 2) to be in 2D. However, there are very few 2D materials in the world. One exception is graphene, which belongs to the 2D point group $6mm$. A representative 2D superlattice material is the graphene nanomesh [12] shown in Fig. 5(a), which is fabricated by perforating holes in a hexagonal pattern to tune its electronic property. The overall point group $\mathcal{G}$ is also $6mm$. Of course, the overall point group could be adjusted by changing the superlattice pattern. For example, a graphene nanomesh with holes distributed in a square pattern [40] has an overall point group $\mathcal{G}$ of $2mm$.

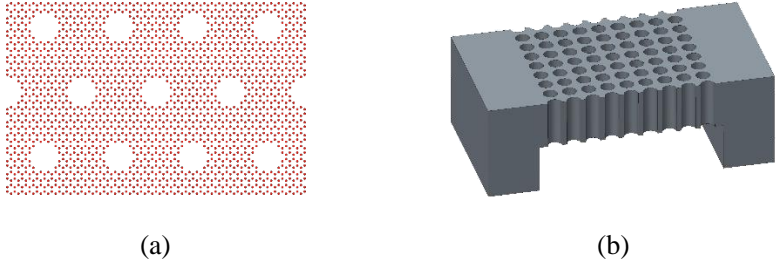

(a)                (b)

**FIGURE 5**. (a) A graphene nanomesh with holes perforated in a hexagonal pattern. (b) A Si-based phononic crystal with a 2D superlattice structure.

Since 2D materials are very rare, most 2D superlattice materials are actually composed of 3D materials, like the Si-based phononic crystal in Fig. 5(b). Even though the superlattice is in 2D, it is better to study the 3D point group of this kind of materials; otherwise, the out-of-plane symmetry property will be missed. In order to use the proposed theory, the 2D topology point group of the superlattice is augmented to 3D by assuming the superlattice unit cell has an infinitesimal out-of-plane lattice parameter. Therefore, the superlattice unit cell in Fig. 5(b) is indeed a tetragonal type instead of a 2D square one. Correspondingly, the Si-based phononic crystal has a topology symmetry $\mathcal{T}$ of $4/mmm$ in 3D rather than $4mm$ in 2D. The overall symmetry $\mathcal{G}$ also depends on the material point group $\bar{\mathcal{M}}$ of the Si, which has an FCC diamond structure with a point group of $m\bar{3}m$. If we consider that the two levels of lattices have the same orientation, the overall symmetry $\mathcal{G}$ of the Si-based phononic crystal would be $4/mmm$, a subgroup of $m\bar{3}m$.

### 4.3  1D Superlattice Materials

One-dimensional (1D) superlattice materials are usually easier to synthesize/fabricate than the 2D or 3D ones. They have been proved to be quite useful and effective for designing phononic/photonic crystals, heat conducting materials, and thermoelectric materials [13, 41, 42]. Three representative examples are introduced to show their symmetry characteristics.

A Si/Ge superlattice nanowire [41] is shown in Fig. 6(a). It is known that both Si and Ge have the diamond cubic crystal lattice, which belongs to the $m\bar{3}m$ point group. In addition, the axial direction of the nanowire is along [001] of the crystal lattice and the lateral facets are the planes {110}, as illustrated in Fig. 6(a). In this case, the topology point group $\mathcal{T}$ of the superlattice is actually the tetragonal group $4/mmm$, which keeps the material type field invariant. Therefore, the overall point group $\mathcal{G} = \mathcal{T} \cap \bar{\mathcal{M}}$ is $4/mmm$ since this is the special case C1. It should be noticed that the topology point group may change if the nanowire is not in such a regular shape.

Figure 6(b) shows another 1D Si/Ge superlattice material [42], which spans through the two directions perpendicular to the 1D superlattice axis. For this class of 1D superlattice materials, the topology point group $\mathcal{T}$ is actually the continuous group $\infty/mm$, because there is no topology boundary in the two in-plane directions. The remaining analysis would be similar to the example in Fig. 6(a) and the overall point group is $4/mmm$.

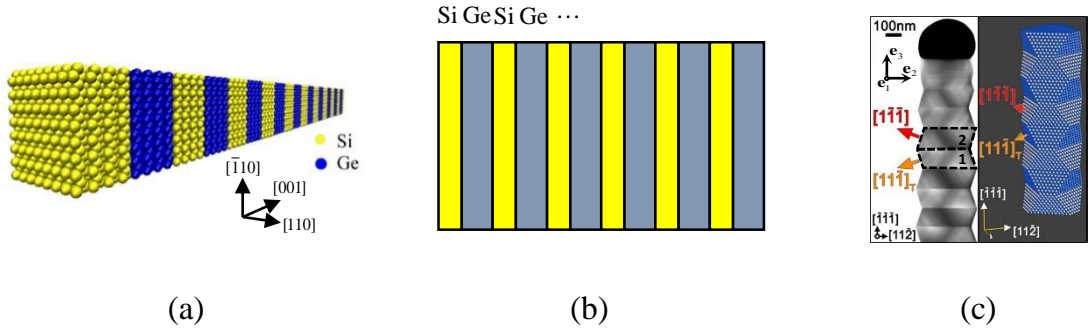

(a) (b) (c)

**FIGURE 6.** (a) Si/Ge superlattice nanowire. (Image reprinted with permission from [41]. ©2012 American Chemical Society) (b) 1D Si/Ge superlattice material. (c) GaAs twinning superlattice. A superlattice unit cell can be divided into 2 components indicated by dash lines. A global coordinate system is established with $\mathbf{e}_1$, $\mathbf{e}_2$, and $\mathbf{e}_3$ as the axes. (Image reprinted with permission from [43]. ©2013 American Chemical Society)

The proposed theory can also be applied to analyze the symmetry of twinning superlattices [43, 44], e.g. the one shown in Fig. 6(c). It is well known that the twin boundary is a reflection plane. However, the overall point group is not easy to determine. The current theory provides a feasible way to determine the overall point group of a twinned superlattice systematically and completely. For example, the GaAs twinning superlattice [43] in Fig. 6(c) has an overall point group $\mathcal{G}$ of $\bar{6}2m$, which is the same as its topology point group $\mathcal{T}$ (see Appendix C). However, the GaAs material only has a point group of $\bar{4}3m$. This indicates that a twinning boundary could create some symmetries that are significantly different from the original single crystal material. Actually this is an example of the special case C3 mentioned in Section 3.3, that the overall symmetry could be identical to the topology symmetry by arranging the material components carefully.

## 4.4 Particles

Rigorously, particles cannot be considered to be superlattice materials, but they are actually the basic building blocks of many superlattice materials from 1D to 3D. Figure 7 shows two kinds of nanoparticles: single crystal nanoparticle and twinned nanoparticle. The symmetry analysis for the single crystal nanoparticle is easy. For example, the truncated Au nanoparticle in Fig. 7(a) has an $m\bar{3}m$ point group for $\bar{\mathcal{M}}$, $\mathcal{T}$, and $\mathcal{G}$. Actually this is why truncated particles (with typical facets) are used to show these crystal structures for demonstration purpose since the topology symmetry directly reflects its material symmetry. Another class of nanoparticles are polycrystalline [45], e.g. single-twinned, multi-twinned, etc. For example, the superlattice unit cell of the GaAs twinned superlattice in Fig. 6(b) is a single-twinned nanoparticle. On the other hand, a multi-twinned Au nanoparticle is shown in Fig. 7(b), which has the five-fold rotation symmetry [46]. In this case, the nanoparticle can be divided into five components, each with the material symmetry of $m\bar{3}m$ (with slightly stretching due to the mismatch) but different crystal orientation. The symmetry analysis follows a procedure similar to Appendix C. The results show that both the topology point group and the global point group are $\bar{5}m$, a non-crystalline point group [35]. This is actually the special case C3 as well, that the symmetry of a superlattice material could be designed to be the same as that of its topology. It should be emphasized that the overall point group of particles could be non-crystalline since the translation order in the superlattice level does not exist anymore.

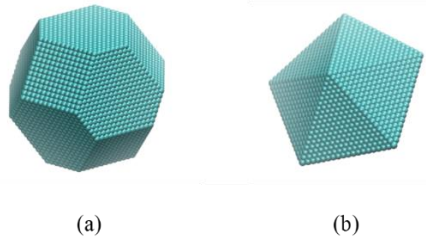

(a)           (b)

**FIGURE 7.** Schematic illustration of Au nanoparticles. The Au crystal has an FCC structure. (a) A truncated octahedron nanoparticle of single crystal. (b) Five-fold decahedral twinned nanoparticle. Images are generated by VMD [47].

## 5 Symmetry of Structural Tensor Field

Other than the point group theory, the material symmetry of superlattice materials can also be characterized by using the structural tensor [30], which has a simpler form and clearer physical meaning. It is known that each point group can be characterized by a single structural tensor. Thus, the structural tensors of the pair of material components in Fig. 3 are studied first. Let us designate $\bar{\mathbf{P}}_{lk}$ as the structural tensor of the material component $\Omega_{lk}$ and $\Omega'_{lk}$ in their local coordinate systems. Generally, $\bar{\mathbf{P}}_{lk}$ is an $n$-th order tensor ($n \geq 1$) and exhibits the following property, as

$$\langle \bar{\mathbf{M}}_{lk} \rangle \bar{\mathbf{P}}_{lk} = \bar{\mathbf{P}}_{lk}, \quad \forall \bar{\mathbf{M}}_{lk} \in \bar{\mathcal{M}}_{lk} \tag{13}$$

where the sign $\langle \square \rangle$ is the form-invariant operator of tensors defined as $\langle \mathbf{A} \rangle \mathbf{B} = A_{il} A_{jm} \cdots A_{kn} B_{lm \cdots n} \mathbf{e}_i \otimes \mathbf{e}_j \otimes \cdots \otimes \mathbf{e}_k$ in general case [30].

In the global coordinate system, the structural tensors of the two material components are derived as

$$\begin{aligned} \mathbf{P}_{lk} &= \langle \mathbf{Q}_{lk}^{\mathrm{T}} \rangle \bar{\mathbf{P}}_{lk} \\ \mathbf{P}'_{lk} &= \langle \mathbf{Q}'^{\mathrm{T}}_{lk} \rangle \bar{\mathbf{P}}_{lk} \end{aligned} \tag{14}$$

Obviously, it is obtained from the first equation in Eq. (14) that

$$\bar{\mathbf{P}}_{lk} = \langle \mathbf{Q}_{lk} \rangle \mathbf{P}_{lk} \tag{15}$$

In addition, based on Eq. (13) and Eq. (15), the structural tensor $\mathbf{P}'_{lk}$ in Eq. (14) can be reformulated as

$$\begin{aligned} \mathbf{P}'_{lk} &= \langle \mathbf{Q}'^{\mathrm{T}}_{lk} \rangle \bar{\mathbf{P}}_{lk} = \langle \mathbf{Q}'^{\mathrm{T}}_{lk} \rangle \langle \bar{\mathbf{M}}_{lk} \rangle \bar{\mathbf{P}}_{lk} = \langle \mathbf{Q}'^{\mathrm{T}}_{lk} \bar{\mathbf{M}}_{lk} \rangle \bar{\mathbf{P}}_{lk} \\ &= \langle \mathbf{Q}'^{\mathrm{T}}_{lk} \bar{\mathbf{M}}_{lk} \rangle \langle \mathbf{Q}_{lk} \rangle \mathbf{P}_{lk} \\ &= \langle \mathbf{Q}'^{\mathrm{T}}_{lk} \bar{\mathbf{M}}_{lk} \mathbf{Q}_{lk} \rangle \mathbf{P}_{lk}, \quad \forall \bar{\mathbf{M}}_{lk} \in \bar{\mathcal{M}}_{lk} \end{aligned} \tag{16}$$

Given that the material symmetry is conserved under a topology symmetry transformation $\mathbf{T}_l \in \mathcal{T}$, it is readily obtained from Eqs. (16) and (9) that

$$\mathbf{P}'_{lk} = \langle \mathbf{T}_l \rangle \mathbf{P}_{lk} \tag{17}$$

Therefore, Eq. (17) indicates that the material symmetry of two material components requires that their structural tensors are form-invariant under $\mathbf{T}_l$.

Consequently, it is concluded that the overall point group $\mathcal{G}$ contains all such topology symmetry transformation $\mathbf{T}_l \in \mathcal{T}$ which conserves the form-invariant of structural tensors for all pairs of material components it transforms. Hence, equivalent to Eq. (11), the point group of superlattice materials can also be determined by

$$\mathcal{G} := \left\{ \mathbf{T}_l \in \mathcal{T} \mid \mathbf{P}'_{lk} = \langle \mathbf{T}_l \rangle \mathbf{P}_{lk} \text{ for } \forall l \leq n_T, \forall k \leq n_l \right\} \tag{18}$$

This structural tensor based definition in Eq. (18) has a simpler form and clearer physical meaning than the point group definition in Eq. (11). Even though, Eq. (11) is still much easier to use since the manipulation of structural tensors is quite awkward.

Actually, similar to Eq. (4), another general definition of the point group $\mathcal{G}$ is that

$$\mathcal{G} := \{ \mathbf{G} \in \mathcal{T} \mid \mathbf{P}(\mathbf{GX}) = \langle \mathbf{G} \rangle \mathbf{P}(\mathbf{X}) \text{ for } \forall \mathbf{X} \in \mathscr{B} \} \tag{19}$$

where $\mathbf{P}(\mathbf{X})$ is the structural tensor field of the superlattice material. Equation (19) is more generalized than Eq. (18) since it adopts a continuum field description. In addition, Eq. (19) also indicates that the material symmetry of superlattice materials is characterized by the symmetry of a tensor field, which is an essential difference from single crystals.

## 6 Symmetry of Deformed Superlattice Materials

### 6.1 Discussion on Symmetry Evolution after Deformation

So far, the point group symmetry theory proposed above is only for the undeformed configuration of superlattice materials. It is already known that the symmetry of superlattice materials might change once they deform [23-25]. However, the symmetry evolution is usually unpredictable in most cases. There are at least two reasons for this difficulty.

First, the lattice type is usually hard to predict after deformation. For example, a uniaxial tensile deformation will change a cubic lattice into a tetragonal lattice with symmetry breaking. In this case, the point group of the deformed lattice is still tractable since it is only a subgroup of the point group of the undeformed lattice. However, the symmetry evolution during the reverse deformation process, i.e. from the tetragonal lattice to the cubic lattice, is intractable since the symmetry lifting occurs. In even worse cases, the point group symmetry could be totally different after deformation once the phase transition occurs. Therefore, the point group after deformation [48] is quite difficult to determine completely unless it is a subgroup of the point group of the undeformed lattice [49, 50]. Fortunately, only symmetry breaking may occur in small deformation cases [50, 51], which will be discussed in details later on.

Second, the material symmetry field of superlattice materials at level 2 is usually unpredictable after deformation. The local material symmetry depends on the local deformation field, which is often intractable for the reason explained above. In addition, the local material symmetry field should also satisfy the overall symmetry throughout the superlattice unit cell.

Therefore, it is hard to establish a unified theory to predict the symmetry evolution of a superlattice material after deformation. We will address two basic problems in Sections 6.2 and 6.3, respectively: (1) In which case the topology symmetry transformation is preserved even after deformation? (2) How does the material symmetry change in small deformation cases?

## 6.2 Deformation that Preserves Topology Symmetry

Without loss of generality, we consider a superlattice material shown in Fig. 8, whose topology symmetry after deformation is studied. Given that an affine lattice deformation [51], represented by a constant deformation gradient tensor $\mathbf{F}_L$, is applied to the lattice points in the superlattice level, the superlattice unit cell will deform from an initial configuration $\mathscr{B} = \{\mathbf{X}\}$ to a deformed configuration $\tilde{\mathscr{B}} = \{\chi(\mathbf{X})\}$ [36], as shown in Fig. 8. The tilde symbol indicates quantities after deformation. In this case, the deformation gradient $\mathbf{F}(\mathbf{X})$ can be decomposed into an affine lattice deformation $\mathbf{F}_L$ and a periodic non-affine deformation field $\mathbf{F}_p(\mathbf{X})$, as [36]

$$\mathbf{F}(\mathbf{X}) \equiv \partial \chi / \partial \mathbf{X} = \mathbf{F}_L \mathbf{F}_p(\mathbf{X}) = \mathbf{R}_L \mathbf{U}_L \mathbf{F}_p(\mathbf{X}) \tag{20}$$

where $\mathbf{R}_L$ and $\mathbf{U}_L$ are the lattice rotation and lattice stretch tensors, respectively [36]. Due to the fact that $\mathbf{F}(\mathbf{X}_L) = \mathbf{F}_L$ for all lattice points, the periodic deformation gradient $\mathbf{F}_p$ satisfies

$$\mathbf{F}_p(\mathbf{X}_L) \equiv \mathbf{I}, \quad \text{for } \forall \mathbf{X}_L \in \mathscr{B} \tag{21}$$

where $\mathbf{X}_L$ represents the lattice points at the superlattice level.

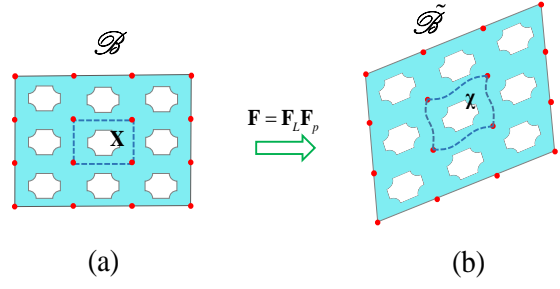

(a)                  (b)

**FIGURE 8**. Schematic illustration for the initial and deformed cellular structures under an affine lattice deformation in the superlattice level. (a) Reference configuration of a rectangular cellular structure. (b) Deformed configuration of the cellular structure. The initial and deformed unit cells are indicated by dashed contour lines.

The symmetry property is studied for the deformed superlattice unit cell. Again, the topology point group $\mathcal{T} = \{\mathbf{T}\}$ is transformed to its conjugacy $\tilde{\mathcal{T}} = \{\tilde{\mathbf{T}}\} = \mathbf{R}_L \mathcal{T} \mathbf{R}_L^T$ in the deformed configuration due to the uniform lattice rotation $\mathbf{R}_L$. Hence under the symmetry transformations $\mathbf{T}$ and $\tilde{\mathbf{T}}$, the reference configuration and deformed configuration are transformed through $\{\mathbf{X}\} = \mathcal{B} \mapsto \mathcal{B}' = \{\mathbf{X}'\}$ and $\{\boldsymbol{\chi}\} = \tilde{\mathcal{B}} \mapsto \tilde{\mathcal{B}}' = \{\boldsymbol{\chi}'\}$, respectively. Then it can be deduced that

$$\begin{aligned} \mathbf{X}' &= \mathbf{T}\mathbf{X} \\ \boldsymbol{\chi}'(\mathbf{X}) &= \tilde{\mathbf{T}}\boldsymbol{\chi}(\mathbf{X}) = \mathbf{R}_L \mathbf{T} \mathbf{R}_L^T \boldsymbol{\chi}(\mathbf{X}) \end{aligned} \tag{22}$$

If the topology symmetry transformation $\tilde{\mathbf{T}}$ is still preserved in the deformed configuration $\tilde{\mathcal{B}} = \{\boldsymbol{\chi}(\mathbf{X})\}$, the symmetry condition is

$$\boldsymbol{\chi}'(\mathbf{X}) = \boldsymbol{\chi}(\mathbf{X}'), \ \forall \mathbf{X} \in \mathcal{B} \tag{23}$$

After substituting Eq. (22) into Eq. (23), it is seen that the deformed configuration should satisfy

$$\mathbf{R}_L \mathbf{T} \mathbf{R}_L^T \boldsymbol{\chi}(\mathbf{X}) = \boldsymbol{\chi}(\mathbf{T}\mathbf{X}), \ \forall \mathbf{X} \in \mathcal{B} \tag{24}$$

Further, taking a first order derivative to $\mathbf{X}$ in Eq. (24) and utilizing Eq. (20) give rise to

$$\mathbf{R}_L^T \mathbf{F}(\mathbf{T}\mathbf{X}) = \mathbf{T} \mathbf{R}_L^T \mathbf{F}(\mathbf{X}) \mathbf{T}^T, \ \forall \mathbf{X} \in \mathcal{B} \tag{25}$$

Equation (25) is the fundamental equation to examine whether the deformation gradient $\mathbf{F}(\mathbf{X})$ preserves the topology symmetry of the superlattice material or not.

The symmetry condition of the deformation gradient $\mathbf{F}(\mathbf{X})$ is equivalent to two separate conditions by using the decomposition in Eq. (20). After substituting Eq. (20) into Eq. (25), we obtain

$$\mathbf{U}_L \mathbf{F}_p(\mathbf{TX}) = \mathbf{TU}_L \mathbf{F}_p(\mathbf{X}) \mathbf{T}^{\mathrm{T}}, \quad \forall \mathbf{X} \in \mathscr{B} \tag{26}$$

By substituting Eq. (21) into Eq. (26) and considering the relation $\mathbf{F}_p(\mathbf{TX}_L) = \mathbf{I}$ implied by Eq. (21), the symmetry of the lattice points $\mathbf{X}_L$ requires that

$$\mathbf{U}_L = \mathbf{TU}_L \mathbf{T}^{\mathrm{T}} \tag{27}$$

Equation (27) indicates that the lattice stretch tensor $\mathbf{U}_L$ should be form-invariant under the symmetry operation $\forall \mathbf{T} \in \mathcal{T}$ if the lattice points are still symmetric after deformation, while the uniform lattice rotation $\mathbf{R}_L$ does not affect the symmetry. The formula in Eq. (27) was derived by Coleman and Noll [50] in another context. Obviously, a uniform dilation deformation, i.e. $\mathbf{U}_L = \lambda \mathbf{I}$ with $\lambda$ as a stretching factor, would not affect the symmetry condition in Eq. (27). The topology symmetry condition of a superlattice material is more complex than Eq. (27) due to the existence of the periodic non-affine deformation $\mathbf{F}_p$. After eliminating $\mathbf{U}_L$ in Eq. (26) by using Eq. (27), the symmetry condition forces the periodic deformation field to satisfy

$$\mathbf{F}_p(\mathbf{TX}) = \mathbf{TF}_p(\mathbf{X}) \mathbf{T}^{\mathrm{T}}, \quad \forall \mathbf{X} \in \mathscr{B} \tag{28}$$

Thus Eq. (28) indicates that the periodic deformation gradient field should be form-invariant under the topology symmetry operation $\forall \mathbf{T} \in \mathcal{T}$.

Further, the symmetry preserving strain field can also be derived. Taking the Green strain tensor field $\mathbf{E}(\mathbf{X}) = [\mathbf{F}^{\mathrm{T}}(\mathbf{X})\mathbf{F}(\mathbf{X}) - \mathbf{I}]/2$ [36] as an example, it satisfies

$$\begin{aligned}
\mathbf{E}(\mathbf{TX}) &= [\mathbf{F}^{\mathrm{T}}(\mathbf{TX})\mathbf{F}(\mathbf{TX}) - \mathbf{I}]/2 \\
&= \mathbf{T}[\mathbf{F}^{\mathrm{T}}(\mathbf{X})\mathbf{F}(\mathbf{X}) - \mathbf{I}]\mathbf{T}^{\mathrm{T}}/2 \\
&= \mathbf{TE}(\mathbf{X})\mathbf{T}^{\mathrm{T}}, \quad \forall \mathbf{X} \in \mathscr{B}
\end{aligned} \tag{29}$$

Figure 9 shows an example [24] of how symmetry breaking occurs when the strain field does not satisfy the symmetry condition in Eq. (29). In this case, Eq. (27) is still valid while Eqs. (28) and (29) are violated. In fact, the strain field does not preserve the $4mm$ topology symmetry of the original configuration since the deformed configuration has a topology point group of $2mm$. This strain-induced symmetry breaking phenomenon has been utilized to tune the phonon propagation behavior in the phononic crystals.

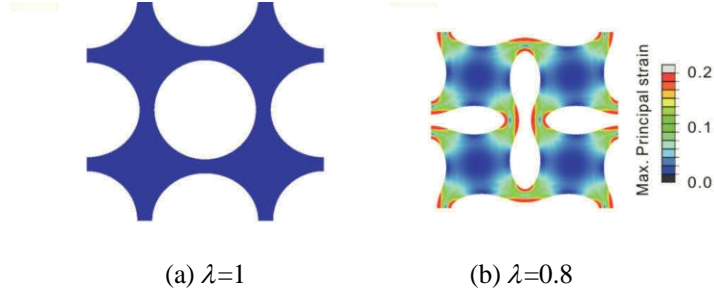

(a) $\lambda=1$  (b) $\lambda=0.8$

**FIGURE 9.** Symmetry breaking of a cellular material induced by deformation. (a) Undeformed configuration. (b) Deformed configuration. $\lambda$ is the uniform stretch factor. (Images reprinted with permission from [24]. ©2013 American Physical Society)

## 6.3  Material Symmetry Breaking in Small Deformation

The material symmetry of the level 2 is studied for superlattice materials. During the deformation process $\mathbf{X} \mapsto \chi(\mathbf{X})$, the material point group field also changes as $\mathcal{M}(\mathbf{X}) \mapsto \tilde{\mathcal{M}}(\chi)$, which is usually hard to determine analytically. According to Eq. (3), The material symmetry of the field $\tilde{\mathcal{M}}(\chi)$ requires that

$$\tilde{\mathcal{M}}(\tilde{\mathbf{G}}\chi) = \tilde{\mathbf{G}}\tilde{\mathcal{M}}(\chi)\tilde{\mathbf{G}}^{\mathrm{T}}, \ \forall \tilde{\mathbf{G}} \in \tilde{\mathcal{G}} \text{ and } \forall \chi \in \tilde{\mathscr{B}} \tag{30}$$

where $\tilde{\mathcal{G}} = \mathbf{R}_L \mathcal{G} \mathbf{R}_L^{\mathrm{T}}$ ($\tilde{\mathcal{G}} \leq \tilde{\mathcal{T}}$) is the overall point group after the affine lattice deformation $\mathbf{F}_L$.

In small deformation cases, $\tilde{\mathcal{M}}(\chi)$ could be determined since only symmetry breaking occurs in the local configuration. The polar decomposition $\mathbf{F}(\mathbf{X}) = \mathbf{R}(\mathbf{X})\mathbf{U}(\mathbf{X})$ is introduced, where $\mathbf{R}(\mathbf{X})$ and $\mathbf{U}(\mathbf{X})$ are the rotation and stretch tensors. For small deformation case, symmetry breaking may occur and the material point group $\tilde{\mathcal{M}}(\chi(\mathbf{X}))$ of the deformed material is [50]

$$\tilde{\mathcal{M}}(\chi(\mathbf{X})) := \{\mathbf{RMR}^{\mathrm{T}} \mid \mathbf{U}(\mathbf{X}) = \mathbf{MU}(\mathbf{X})\mathbf{M}^{\mathrm{T}}, \text{ for } \forall \mathbf{M} \in \mathcal{M} \text{ and } \forall \mathbf{X} \in \mathscr{B}\} \tag{31}$$

On the other hand, the material symmetry evolution is quite complex and almost intractable for large deformation cases, which should be paid careful attention to.

## 7  Remarks on Physical Properties of Superlattice Materials

The physical properties of materials are closely related to their symmetry conditions [21, 22]. Compared to single crystals, the physical properties of superlattice materials exhibit some new features due to their structural and material diversity.

First, the physical properties of superlattice materials depend on the symmetry conditions at different length scales. The analysis of the long-range physical properties of superlattice materials is similar to crystals [52-54]. The point group of superlattice materials should be within the aforementioned 51 point groups, e.g. 39 in 3D and 12 in 2D. Therefore, the basic forms of long-range physical properties, which are characterized by different ranks of tensors $\mathbf{K}$, could be formulated once the point group is determined [21, 22]. Typical examples of the physical properties are density and heat capacity (rank 0), dielectricity and conductivity (rank 2), elasticity (rank 4), just name a few. The point group symmetry renders the physical property to satisfy [21]

$$\langle \mathbf{G} \rangle \mathbf{K} = \mathbf{K}, \forall \mathbf{G} \in \mathcal{G} \tag{32}$$

The point group would usually simplify the forms of physical properties to a large extent; however, this simplification is dangerous sometimes. Note that the point group symmetry only determines the long-range physical properties, which means that the characteristic length $l$ is much larger than the superlattice unit cell size $a$. In contrast, short-range physical properties ($l \ll a$) are mainly dependent on the local material point group symmetry. The intermediate-range physical properties are more complex, which are related to the interactions between the global symmetry and local material symmetry.

Another feature of superlattice materials is that their physical properties can be designed and tuned in favor of the required performance. The unit cell size, topology, material type, and material distribution in superlattice materials can all be designed and optimized to achieve the desired physical properties [5-16], which is an advantage that crystals and other homogeneous materials do not exhibit. Particularly, this property-by-design methodology is growing in an unprecedented pace recently owing to the eruption of new synthesis and manufacturing technologies from nano- to macro-scale. Recent literature also reports that the symmetry of superlattice materials can be tuned via deformation [23-25, 27] to achieve tunable or controllable physical properties, which further broadens the applications of superlattice materials.

In brief summary, the physical properties of superlattice materials strongly depend on their symmetry properties at multiple length scales. Utilizing the symmetry conditions will greatly facilitate superlattice material design with multifunctional usage and exceptional properties.

## 8 Conclusions

Point group symmetry is one of the most important and fundamental properties of materials, which is related to their physical properties and useful for materials modeling. However, the symmetry of superlattice materials is more complicated than the conventional crystal symmetry since the superlattice materials require both topology symmetry and material symmetry in one superlattice unit cell. To address this problem, a unified theoretical framework is established to describe and determine the overall point group of superlattice materials. Current work reveals that the point group symmetry of superlattice materials can be described by the invariant (or form-invariant) of a material type field and a material point group field, with the latter equivalent to the structural tensor field. This is significantly different from the symmetry of single crystals, which only requires the invariant of the material type field. The proposed theory is explained and applied to a variety of examples to show the symmetry properties of superlattice materials ranging from 1D to 3D. In addition, the point group symmetry evolution for deformed superlattice materials is also discussed with an emphasis on the deformation induced symmetry breaking phenomena for small strain cases. The proposed theory will provide theoretical foundation for studying the physical properties of superlattice materials and tuning the physical properties via symmetry breaking.

**Authors' contributions.** P.Z. carried out the research and wrote the manuscript with the assistance and supervision of A.C.T.

**Conflict of interests.** We have no competing interests.

**Appendix A. Symmetry Transformations of 3D Material Point Groups**

There are totally 32 crystal point groups and 7 continuous point groups for 3D materials [35]. Each crystal point group contains a finite number of symmetry transformations, whereas the continuous point groups are non-compact. In order to describe these symmetry transformations, a Cartesian coordinate system is established, which has three orthogonal axes $\mathbf{e}_1$, $\mathbf{e}_2$, and $\mathbf{e}_3$. Each crystal class has three preferred lattice vectors $\mathbf{a}_1$, $\mathbf{a}_2$, and $\mathbf{a}_3$ according to the convention of crystallography [35]. The default material orientations are introduced first, which follow the notations used in [38, 39]. (1) Triclinic lattices can be arbitrarily oriented. (2) The lattice vector $\mathbf{a}_1$ is parallel to the axis $\mathbf{e}_1$ for monoclinic lattices. (3) For the rhombic, tetragonal, and cubic systems, $\mathbf{a}_1$, $\mathbf{a}_2$, and $\mathbf{a}_3$ are parallel to the axes $\mathbf{e}_1$, $\mathbf{e}_2$, and $\mathbf{e}_3$, respectively. (4) For the hexagonal system, $\mathbf{a}_1$ and $\mathbf{a}_3$ are parallel to the axes $\mathbf{e}_1$ and $\mathbf{e}_3$, respectively. Note that the trigonal crystal system is classified into the hexagonal system here [38]. The detailed symmetry transformations of 32 crystal point groups are listed in [38, 39].

Besides the 32 crystal point groups, there are two isotropy groups ($\infty\infty m$ and $\infty\infty$) and five transverse isotropy groups ($\infty$, $\infty m$, $\infty/m$, $\infty 2$, and $\infty/mm$) in 3D [21, 35]. The group $\infty\infty m$ is equal to the 3D orthogonal group $\mathcal{O}(3)$, while the group $\infty\infty$ is equal to the 3D proper orthogonal group $\mathcal{O}^+(3)$. On the other hand, all transverse isotropy groups have a preferred rotation axis along $\mathbf{e}_3$. The rotation transformation is represented by a continuous function, as

$$\mathbf{M}_\theta = \begin{bmatrix} \cos\theta & \sin\theta & 0 \\ -\sin\theta & \cos\theta & 0 \\ 0 & 0 & 1 \end{bmatrix} \tag{33}$$

where $0 \leq \theta \leq 2\pi$. All the symmetry transformations of transverse isotropic groups [38] can be generated by $\mathbf{M}_\theta$ and the matrices $\mathbf{R}_1 = \mathrm{diag}(-1,1,1)$, $\mathbf{R}_3 = \mathrm{diag}(1,1,-1)$, and $\mathbf{D}_2 = \mathrm{diag}(-1,1,-1)$. All 3D continuous point groups are listed in Table 1.

The 2D material point groups [35] could be degenerated from the 3D groups, which will not be introduced in details.

**Table 1**. Symmetry transformations of continuous point groups in 3D

| Class | Symmetry Transformations |
|---|---|
| $\infty\infty m$ | $\mathcal{O}(3)$ |
| $\infty\infty$ | $\mathcal{O}^+(3)$ |
| $\infty$ | $\mathbf{M}_\theta$ |
| $\infty m$ | $\mathbf{M}_\theta$, $\mathbf{R}_1$, $\mathbf{M}_\theta \mathbf{R}_1$ |
| $\infty/m$ | $\mathbf{M}_\theta$, $\mathbf{R}_3$, $\mathbf{M}_\theta \mathbf{R}_3$ |
| $\infty 2$ | $\mathbf{M}_\theta$, $\mathbf{D}_2$, $\mathbf{M}_\theta \mathbf{D}_2$ |
| $\infty/mm$ | $\mathbf{M}_\theta$, $\mathbf{R}_1$, $\mathbf{R}_3$, $\mathbf{D}_2$, $\mathbf{M}_\theta \mathbf{R}_1$, $\mathbf{M}_\theta \mathbf{R}_3$, $\mathbf{M}_\theta \mathbf{D}_2$ |

**Appendix B. Point Group Symmetry of Nanocrystal Superlattice**

The overall point group symmetry of the nanocrystal superlattice shown in Fig. 1(b) is determined in the following procedure. It is known that both $\mathcal{T}$ and $\bar{\mathcal{M}}$ belong to the group $m\bar{3}m$. Therefore, it is found that

$$\mathcal{T} = \bar{\mathcal{M}} = \{\mathbf{I}, \mathbf{C}, \mathbf{R}_1, \mathbf{R}_2, \mathbf{R}_3, \cdots\}_{48} \tag{34}$$

where the subscript '48' indicates the order of the point group. All the 48 symmetry transformations are outlined in Refs. [38, 39] with the same notation. The overall point group is determined by $\mathcal{G} = \mathcal{T} \cap \mathbf{Q}^T \bar{\mathcal{M}} \mathbf{Q}$ with $\mathbf{Q}$ shown in Eq. (12). Hence, we obtain

$$\begin{aligned}\mathcal{G} &= \mathcal{T} \cap \mathbf{Q}^T \bar{\mathcal{M}} \mathbf{Q} \\ &= \{\mathbf{I}, \mathbf{T}_3, \mathbf{R}_3, \mathbf{D}_3\mathbf{T}_3, \mathbf{C}, \mathbf{CT}_3, \mathbf{CR}_3, \mathbf{CD}_3\mathbf{T}_3\} \\ &= \hat{\mathbf{Q}} \bar{\mathcal{M}}_{mmm} \hat{\mathbf{Q}}^T \end{aligned} \tag{35}$$

where $\bar{\mathcal{M}}_{mmm}$ represents the $mmm$ point group listed in Refs. [38, 39] and $\hat{\mathbf{Q}}$ is determined as

$$\hat{\mathbf{Q}} = \begin{bmatrix} \frac{1}{\sqrt{2}} & \frac{1}{\sqrt{2}} & 0 \\ \frac{-1}{\sqrt{2}} & \frac{1}{\sqrt{2}} & 0 \\ 0 & 0 & 1 \end{bmatrix} \tag{36}$$

Equation (35) indicates that the overall point group $\mathcal{G}$ is a conjugate group of $mmm$. In addition, the three reflection planes are $(110)$, $(1\bar{1}0)$, and $(001)$ of the superlattice according to Eq. (36).

**Appendix C. Point Group Symmetry of GaAs Twinning Superlattice**

The overall point group of the GaAs twinning superlattice in Fig. 6(c) is explained briefly. The local coordinate systems on component 1 and 2 of the superlattice unit cell can be transformed from the global coordinate system by $\mathbf{Q}_1$ and $\mathbf{Q}_2$, as

$$\mathbf{Q}_1 = \begin{bmatrix} \frac{-1}{\sqrt{2}} & \frac{1}{\sqrt{6}} & \frac{-1}{\sqrt{3}} \\ \frac{1}{\sqrt{2}} & \frac{1}{\sqrt{6}} & \frac{-1}{\sqrt{3}} \\ 0 & \frac{-2}{\sqrt{6}} & \frac{-1}{\sqrt{3}} \end{bmatrix}, \quad \mathbf{Q}_2 = \begin{bmatrix} \frac{1}{\sqrt{2}} & \frac{1}{\sqrt{6}} & \frac{1}{\sqrt{3}} \\ \frac{-1}{\sqrt{2}} & \frac{1}{\sqrt{6}} & \frac{1}{\sqrt{3}} \\ 0 & \frac{-2}{\sqrt{6}} & \frac{1}{\sqrt{3}} \end{bmatrix} \tag{37}$$

For this example, the symmetry point group $\mathcal{T}$ and $\bar{\mathcal{M}}$ are given by [38, 39]

$$\mathcal{T} = \{\mathbf{I}, \mathbf{S}_1, \mathbf{S}_2, \cdots\}_{12}, \quad \bar{\mathcal{M}} = \{\mathbf{I}, \mathbf{D}_1, \mathbf{D}_2, \cdots\}_{24} \tag{38}$$

It can be verified that the transformations $\{\mathbf{I}, \mathbf{S}_1, \mathbf{S}_2, \mathbf{R}_1, \mathbf{R}_1\mathbf{S}_1, \mathbf{R}_1\mathbf{S}_2\} \subset \mathcal{T}$ will transform a material component into itself and these 6 transformations satisfy $\mathbf{T} \in \mathbf{Q}_1^T \bar{\mathcal{M}} \mathbf{Q}_1$ and $\mathbf{T} \in \mathbf{Q}_2^T \bar{\mathcal{M}} \mathbf{Q}_2$. In addition, the other six transformations $\{\mathbf{R}_3, \mathbf{R}_3\mathbf{S}_1, \mathbf{R}_3\mathbf{S}_2, \mathbf{D}_2, \mathbf{D}_2\mathbf{S}_1, \mathbf{D}_2\mathbf{S}_2\} \subset \mathcal{T}$ will transform the two material components into each other and they satisfy $\mathbf{T} \in \mathbf{Q}_2^T \bar{\mathcal{M}} \mathbf{Q}_1$. Therefore, the overall point group is identical to the topology symmetry point group, i.e. $\mathcal{G} = \mathcal{T}$, because the condition for the special case C3 is satisfied.